\documentclass[fleqn,twoside]{article}
\usepackage{graphicx}
\usepackage{espcrc2}

\title{Dynamical overlap fermions, results with HMC algorithm}

\author{
Z. Fodor\address[BUW]{Department of Physics, University of Wuppertal,
        Gaussstrasse 20, D-42097  Wuppertal, Germany} 
\address{Institute for Theoretical Physics, E\"otv\"os University,
P\'azm\'any P. 1/A, H-1117 Budapest, Hungary},
S. D. Katz\addressmark[BUW]\thanks{On leave from E\"otv\"os University.}, 
K. K. Szabo\addressmark[BUW]
}

\begin{document}

\begin{abstract} 
We present results of a hybrid Monte-Carlo algorithm for dynamical, $n_f=2$,
four-dimensional QCD with overlap fermions.  The fermionic force requires careful treatment, when changing 
topological sectors. The pion mass dependence of the 
topological susceptibility is studied on $6^4$ and $12\cdot 6^3$ lattices. The results are transformed into physical units. 
\end{abstract}

\maketitle

\section{HMC FOR THE OVERLAP}
It is a great numerical challenge to make dynamical QCD simulations with Ginsparg-Wilson fermions, which have exact chiral
symmetry at finite lattice spacing. We present an exploratory study for overlap fermions, 
which have the Dirac
operator:
\[D=m_0(1+\gamma_5 {\rm sgn}(H_W)),\]
where $H_W$ is the hermitian Wilson-operator with $-m_0$ mass \cite{Neuberger:1997fp}. The bare mass is introduced by:
\[
D(m)=m+\left(1-m/(2m_0)\right) D.
\]

We use the Zolotarev rational series to approximate the sign function:
\[{\rm sgn}(x) \approx \epsilon_n(x)
=x(x^2+c_{2n})\sum_{l=1}^n {b_l \over x^2+c_{2l-1}},\]
for $b_l,c_l$ see \cite{vandenEshof:2002ms}.  
The evaluation of the inverse of the
shifted Wilson-matrices 
is done by multishift conjugate gradient algorithm \cite{Frommer:1995ik}. To speed up the inversions one can project out the
low-lying eigenmodes of the Wilson kernel:
\[ 
{\rm sgn} (H_W) \approx \sum_i {\rm sgn}(\lambda_i) 
|\lambda_i \rangle \langle \lambda_i | +T \epsilon _n (H_W),  
\]
where $H_W|\lambda_i\rangle = \lambda_i |\lambda_i \rangle$ and $T=1-\sum_i |\lambda_i \rangle \langle \lambda_i |$ 
projects to the orthogonal subspace.

We use the HMC algorithm to generate dynamical configurations for $n_f=2$ flavors. Thus we need the expression of the
microcanonical energy:
\[ 
\mathcal{H}=
\frac{1}{2}\langle P,P\rangle+S_{\rm g}[U]+\phi^\dagger (D^\dagger D)^{-1} \phi,
\] 
which governs a classical motion in the space of link variables. $\phi$ is the pseudofermion field, whereas
$P$ is the canonical momenta, which is randomized at the beginning of each trajectory. For the classical
motion we need the gauge derivative of pseudofermionic action, namely the fermion force.

For the fermion force one needs to invert the overlap operator on the $\phi$ field:
\[{\rm force}=-\psi^\dagger {d D^\dagger D\over dU} \psi, \quad {\rm where} \quad \psi=(D^\dagger D)^{-1}
\phi,\]
thus one ends up with a nested inversion
procedure, which is the most time consuming part of the algorithm. The gauge derivative of the rational
approximation is straightforwardly calculated, furthermore it is easy to include the contribution of the
projected eigenmodes exactly in the force \cite{Liu:1998hj}. Then one can solve the equations of motion with a reversible
and area conserving integration procedure (usually it is a simple leapfrog). 
\section{FERMION FORCE}
With the naive application of the above results one faces a further problem. The
pseudofermionic action has a dicontinuity across the surface, where $H_W$ matrix has a zeromode, which means a
presence of a Dirac-delta type singularity in the force.
A finite
stepsize integration of the equations of motion practically never notice it. This means that it 
drastically violates energy conservation.
Finally one ends up with a very bad acceptance ratio, since the accept/reject
step depends exponentially on the energy-conservation violation. We have observed that already on a $6^4$
lattice there were no accepted configurations at all.

These eigenvalue crossings are easily connected with the topological sector change, since topological charge 
$Q$ changes whenever a $\lambda_i$ crosses zero:
\[
\Delta Q=\frac{1}{2m_0}\Delta [ {\rm Tr}(\gamma_5 D) ] = \frac{1}{2}\Delta [ 
\sum_i {\rm sgn}(\lambda_i) ]
= \pm 1.
\]

We cure the problem by considering a classical trajectory, when it reaches the zero
eigenvalue surface. Let the jump in the action on the surface be $\Delta S$, whereas the normalvector of the
surface is $N=\langle \lambda | dH_W/dU |\lambda \rangle$. If the momentum is not large enough to climb the
jump ($P^2 < 2 \Delta S$),  
the trajectory will be reflected off from the surface, its new momentum will  be:
\[
P'=P-2N\langle N,P \rangle.
\]
If there is enough kinetic energy ($P^2 > 2 \Delta S$), then the system will move into the other topological sector with a new
momentum:
\[
P'=P-N\langle N, P \rangle + N \langle N, P \rangle \sqrt{1-2\Delta S/\langle N, P \rangle ^2} .
\] 

One should modify the standard leapfrog algorithm to accomodate the above momentum change. The eigenvalue
crossing can happen only in the half step, when the gauge fields are updated. We propose to replace this step with three other
ones (corrected leapfrog step):\\ 
1. Evolve the gauge fields just to the eigenvalue surface ($\tau_1$ is the time, which is needed for this).\\
2. Decide whether reflection or refraction happened, and modifiy the momentum according to the above rules.\\
3. Update the gauge fields in the remaining time ($\tau/2-\tau_1$) with the new momenta.\\
The algorithm is reversible, preserves area and conserves energy only upto $O(\tau_1)$. 
It is an important goal of future algorithm developments to improve on this behaviour.

The stepsize should be small enough to keep track the evolution of eigenvalues, and to avoid of
the problem of having 2,3 or more crossing eigenvalues in one microcanoncial step. 

\begin{figure}
\includegraphics*[width=8cm,bb=18 202 575 717]{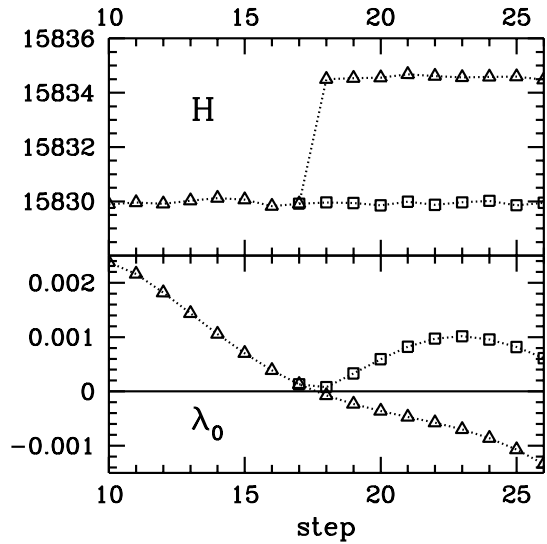}
\vspace{-1.3cm}
\caption{Comparison of the uncorrected (triangles) and corrected (boxes) leapfrogs. 
Upper part is the energy, lower part is the lowest
mode of $H_W$.}
\label{fig:ref}
\end{figure}

Fig. \ref{fig:ref} compares the evolution of the energy and lowest $\lambda$ for the usual and for the
corrected leapfrog. In the uncorrected case there is a huge energy jump at the crossing, whereas in the modified case a
reflection happens, and $\mathcal{H}$ is much better conserved. 

\section{NUMERICAL RESULTS}
We have checked our code by a brute force approach on $2^4$ and $4^4$ lattices: we have weighted a quenched ensemble with the 
exactly calculated determinant. A complete agreement was found. On $4\cdot 6^3$ lattices there is a sharp
increase in the Polyakov loop at $\beta=5.7$, this value of the coupling was used for measuring the topology on $6^4$
lattices.
The negative quark mass was set to $m_0=1.6$, the bare fermion mass was in the range $m=0.1..1.15$, the
stepsize was $\tau=0.025$ in average. At each bare mass roughly 800 trajectories were generated.  

\begin{figure}
\vspace{-1.0cm}
\caption{Topology on $N_s=6$ lattices. See text.}
\vspace{-0.5cm}
\includegraphics*[width=7cm,bb=36 184 216 707]{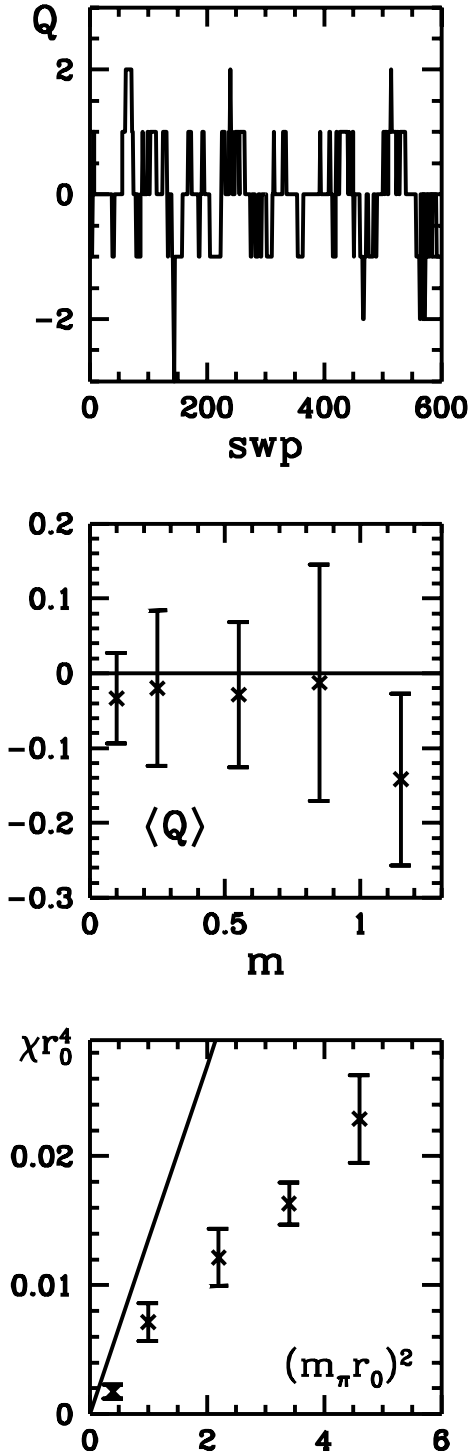}
\label{fig:ts}
\end{figure}
The results are plotted on Fig. \ref{fig:ts}. 
The upper panel shows the charge history. The average topological charge is consistent
with zero for the total mass range (middle panel). 
$12\cdot 6^3$ lattices were used to fix the scale using $r_0$ from Wilson-loops. The result is $a \sim
0.25$ fm for small masses. Pion masses were also measured and $m_\pi^2=Am$ with $A\approx 1$ is found in lattice units.  
Using the scale and the pion mass, it is possible to get the topological susceptibility in physical units
(lower panel of Fig. \ref{fig:ts}). 
$\chi(m)$ tends to zero for small quark masses. One can compare these results with the continuum 
expectation in the chiral limit (solid line of the figure):
\[
\lim_{m \to 0} \chi(m) = \lim_{m\to 0}\frac{\langle Q^2 \rangle
}{V}=\frac{f_\pi^2m_\pi^2}{2n_f}. 
\]

Caution is needed, when interpreting the results. There are obvious problems such as
small volume and rough lattice.
Furthermore there can be doubts about the locality of the theory at these rather large couplings \cite{Golterman:2003qe}. 

A detailed description of the algorithm can be found in \cite{Fodor:2003bh}, whereas for an independent study see \cite{Cundy:2004xf}.

{\bf Acknowledgements:}
We thank Tam\'as G. Kov\'acs for useful discussions.
This work was partially supported by Hungarian Scientific
grants 
OTKA-T37615/\-T34980/\-T29803/\-TS44839/\-T46925. 
The simulations were carried out on the  
E\"otv\"os Univ., Inst. Theor. Phys. 330 P4 node parallel PC cluster.


\begin{thebibliography}{9}
\bibitem{Neuberger:1997fp}
H.~Neuberger,
Phys.\ Lett.\ B {\bf 417} (1998) 141


\bibitem{vandenEshof:2002ms}
J.~van den Eshof et al.,
Comput.\ Phys.\ Commun.\  {\bf 146} (2002) 203
\bibitem{Frommer:1995ik}
A.~Frommer et al.,
Int.\ J.\ Mod.\ Phys.\ C {\bf 6} (1995) 627
\bibitem{Liu:1998hj}
C.~Liu,
Nucl.\ Phys.\ B {\bf 554} (1999) 313;
R.~Narayanan and H.~Neuberger,
Phys.\ Rev.\ D {\bf 62} (2000) 074504
\bibitem{Golterman:2003qe}
M.~Golterman and Y.~Shamir,
Phys.\ Rev.\ D {\bf 68} (2003) 074501
\bibitem{Fodor:2003bh}
Z.~Fodor, S.~D.~Katz and K.~K.~Szabo,
JHEP {\bf 0408}, 003 (2004)

\bibitem{Cundy:2004xf}
N.~Cundy et al.
arXiv:hep-lat/0405003;
N.~Cundy et al.
arXiv:hep-lat/0409029.

\end{thebibliography}
\end{document}